\documentclass[%
aip,%
amsmath,amssymb,floatfix,
reprint,%
onecolumn,
jcp,%
longbibliography,
]{revtex4-1}

\usepackage{amsmath,amsthm,latexsym,amssymb,amsfonts,epsfig}

\usepackage[english]{babel}
\usepackage[utf8]{inputenc}			
\usepackage{graphicx, texdraw} 
\usepackage{color}
\usepackage{bm}

\usepackage{mathtools}
\usepackage{mathrsfs} 

\usepackage[left= 2.cm, right = 2cm, top= 2.2cm, bottom = 2.2cm]{geometry}  

\usepackage{enumerate}

\usepackage[colorlinks=true,citecolor=OliveGreen, linkcolor=blue]{hyperref}

\usepackage{csquotes} 

\usepackage{color}
\usepackage[svgnames,dvipsnames]{xcolor}

\usepackage{amsmath}
\usepackage{amsfonts}
\usepackage{amssymb}

\usepackage{epstopdf}

\usepackage{type1ec}

\newcommand{\vect}[1]{\boldsymbol{#1}}

\newcommand{\bNabla}{\boldsymbol{\nabla}}

\newcommand{\bNablaParallel}{\bNabla_\parallel}

\newcommand{\Vpcom}{V_{\mathrm{P}}}

\newcommand{\sig}{\vect{\sigma}}

\newcommand{\uu}{\vect{u}}
\newcommand{\xS}{\vect{x}_{\mathrm{S}}}

\newcommand{\vv}{\vect{v}}

\newcommand{\kS}{\kappa_\mathrm{S}}
\newcommand{\kA}{\kappa_\mathrm{A}}
\newcommand{\kB}{\kappa_\mathrm{B}}

\newcommand{\LambdaB}{\Lambda_\mathrm{B}}
\newcommand{\LambdaS}{\Lambda_\mathrm{S}}

\newcommand{\X}{\vect{x}}

\newcommand{\Intd}{\mathrm{d}}

\providecommand*{\ez}{\bm{e}_{z}}

\providecommand*{\pp}[3][]{\frac{\partial^{#1}#2}{\partial #3^{#1}}}

\newcommand{\qmbox}[1]{\quad\mbox{#1}\quad}

\providecommand*{\rmd}{\mathrm{d}}

\newcommand{\arctanh}{\operatorname{arctanh}}

\allowdisplaybreaks

\begin{document}
\title{A reciprocal theorem for the prediction of the normal force induced on a particle translating parallel to an elastic membrane}

\author{Abdallah Daddi-Moussa-Ider}
\email{ider@thphy.uni-duesseldorf.de}
\thanks{ADMI and BR contributed equally to this work.}

\affiliation
{Institut f\"{u}r Theoretische Physik II: Weiche Materie, Heinrich-Heine-Universit\"{a}t D\"{u}sseldorf, Universit\"{a}tsstra\ss e 1, D\"{u}sseldorf 40225, Germany}

\affiliation
{Biofluid Simulation and Modeling, Theoretische Physik, Universit\"at Bayreuth, Universit\"{a}tsstra{\ss}e 30, Bayreuth 95440, Germany}

\author{Bhargav Rallabandi}

\affiliation
{Department of Mechanical and Aerospace Engineering, Princeton University, Princeton, NJ 08544, United  States}

\author{Stephan Gekle}
\affiliation
{Biofluid Simulation and Modeling, Theoretische Physik, Universit\"at Bayreuth, Universit\"{a}tsstra{\ss}e 30, Bayreuth 95440, Germany}

\author{Howard A. Stone}
\email{hastone@princeton.edu}
\affiliation
{Department of Mechanical and Aerospace Engineering, Princeton University, Princeton, NJ 08544, United  States}

\begin{abstract}

When an elastic object is dragged through a viscous fluid tangent to a rigid boundary, it experiences a lift force perpendicular to its direction of motion. 
An analogous lift mechanism occurs when a rigid symmetric object translates parallel to an elastic interface or a soft substrate.
The induced lift force is attributed to an elastohydrodynamic coupling that arises from the breaking of the flow reversal symmetry induced by the elastic deformation of the translating object or the interface. 
Here we derive explicit analytical expressions for the quasi-steady state lift force exerted on a rigid spherical particle translating parallel to a finite-sized membrane exhibiting a resistance toward both shear and bending.
Our analytical approach proceeds through the application of the Lorentz reciprocal theorem so as to obtain the solution of the flow problem using a perturbation technique  for small deformations of the membrane.
We find that the shear-related contribution to the normal force leads to an attractive interaction between the particle and the membrane.
This emerging attractive force decreases quadratically with the system size to eventually vanish in the limit of an infinitely-extended membrane.
In contrast, membrane bending leads to a repulsive interaction whose effect becomes more pronounced upon increasing the system size, where the lift force is found to diverge logarithmically for an infinitely-large membrane.
The unphysical divergence of the bending-induced lift force can be rendered finite by regularizing the solution with a cut-off length beyond which the bending forces become subdominant to an external body force.

\end{abstract}

\date{\today}
\maketitle

\section{Introduction}

The coupling between soft boundaries and viscous flows plays an important role in many physical phenomena and finds applications in a large variety of fields in engineering and science~\cite{dowson14}.
Notable examples include the emergence of surface-tension-driven coalescence of flexible structures~\cite{duprat11}, the deformation of slender elastic filaments during sedimentation~\cite{stone15}, the elastohydrodynamic wake generated in a thin lubricated elastic sheet~\cite{ledesma16, arutkin17, domino18}, the formation of biofilm streamers in microchannels~\cite{rusconi10, rusconi11, drescher13, kim14}, 
the propulsion of elastica in a viscous fluid~\cite{wiggins98, yu06}, and the elastocapillary soft leveling of thin viscous films on elastic substrates~\cite{rivetti17}.
Elastohydrodynamic effects may have significant consequences in a wide range of biological and physiological processes, ranging from the rheology of a suspension of red blood cells in microcapillaries~\cite{secomb86, pozrikidis05axi, mcwhirter09, Freund_2014, secomb17, guckenberger18} to the lubrication of synovial joints in the limbs~\cite{walker68, dowson86, jin05}.

In low-Reynolds-number hydrodynamics, the motion of suspended particles is described by the linear Stokes equations~\cite{happel12}, where the viscous forces are much larger than the inertial forces. 
Because of the long-range nature of the hydrodynamic interactions, the motion of suspended particles in a viscous fluid is strongly altered by confining interfaces.
As an example, the reversibility of the Stokes equations implies that no lift force is exerted on a rigid symmetric object, such as a sphere or a circular cylinder, that translates parallel to a flat hard wall~\cite{goldman67a, goldman67b}.
However, this reversibility can be broken by introducing nonlinear effects due to inertia \cite{saffman65, drew88, dandy90}, viscoelasticity of the surrounding fluid \cite{hu99, patankar01, pandey16, putignano17}, or the elastic nature of either or both of the translating object and the interface.
For instance, a capsule that is enclosed by an elastic membrane in a wall-bounded shear flow experiences a net non-inertial lateral migration in which the lift velocity increases with the shear rate and decreases with distance from the wall~\cite{sukumaran01, abkarian02, callens08}.

Theoretically, the elastohydrodynamic-induced lift force has been addressed thoroughly in the lubrication limit~\cite{sekimoto93, skotheim04, skotheim05, yin05, snoeijer13, wang15, wang17, wang17SM}, finding that there exists an optimal combination of geometric and material parameters that maximizes the lift force.
Earlier research considered the elastohydrodynamic collision of two spheres via asymptotic analysis \cite{davis86, serayssol86}, and more recently the motion of two elastic bodies at relative speed~\cite{snoeijer13}, the lift force experienced by a small sphere translating and rotating near a soft wall \cite{urzay07, urzay10}, and the lift force induced between polymer-bearing surfaces~\cite{sekimoto93}.
Using a local linear pressure–displacement model for the deformable wall, the transient behavior has also been studied~\cite{weekley06}.
The influence of a deformable substrate on the dynamics of a fluid vesicle moving in its vicinity has been numerically studied, finding that the optimal elastic modulus for the lift force lies within the physiological range~\cite{beaucourt04lift}.
Moreover, it has been shown that reciprocal motion near a deformable interface can circumvent Purcell's scallop theorem~\cite{purcell77} and lead to a net propulsion of swimming microorganisms in low-Reynolds-number locomotion~\cite{trouilloud08}.

More recently, the motion of a negatively buoyant cylinder in the vicinity of an inclined thin compressible elastic wall has been investigated using elastohydrodynamic lubrication theory~\cite{salez15}, showing that different scenarios of motion occur, that relate sedimentation, sliding and spinning motion modes.
Corresponding experiments that have been carried out near a soft incline~\cite{saintyves16} have reported that the translating cylinder further undergoes a spontaneous steady-state rotation.
This behavior has been explained theoretically using a higher-order asymptotic analysis in the lubrication limit~\cite{rallabandi17}.
Meanwhile, the normal displacement of a spherical particle sedimenting under gravity along an elastic membrane has been measured experimentally~\cite{rallabandi18} where good agreement has been obtained with an analytical model based on lubrication theory.
It has been suggested that the observed lift effect can be utilized in the design of size-sorting processes and separation devices.

The slow motion of a spherical solid particle moving near a planar elastic membrane possessing a resistance to shear and bending has been investigated theoretically using a far-field model~\cite{daddi16,daddi17,daddi18epje}.
It has been demonstrated that the elastic nature of the membrane endows the system with memory and leads to a long-lasting anomalous subdiffusive behavior on nearby particles~\cite{daddi16}.
Further theoretical investigations have been carried our near membranes with curved geometries~\cite{daddi17b,daddi17d}, finding that shear usually manifests itself in a more pronounced way compared to bending.
However, the latter studies were limited to the effect of the membrane on the drag force and have not examined the lift force  arising from the nonlinear nature of the elastohydrodynamic problem.
The goal of this paper is to quantify this lift effect and derive explicit analytical expressions for the induced nonlinear normal force.
We find that the lift force is repulsive due to bending while the shear-related contribution to the normal force is found to have an opposite effect.
The latter, however, decays quadratically with increasing system size and vanishes for an infinitely-extended membrane.

In the remainder of this paper, we introduce in Sec.~\ref{sec:theoreticalPredictions} the elastohydrodynamic problem of a solid sphere translating tangent to an elastic membrane and state the governing equations of fluid motion in addition to the underlying boundary conditions.
We then present in Sec.~\ref{sec:reciprocalTheorem} the reciprocal theorem for Stokes flow and derive a general formula for the nonlinear normal force that is induced due to an arbitrary velocity distribution prescribed for a given reference configuration of the membrane.
The calculations of the bending- and shear-related contributions to the normal force are detailed in Sec.~\ref{sec:calculations} where analytical expressions are obtained.
The regularization solution is discussed in Sec.~\ref{sec:regularization}.
Concluding remarks are contained in Sec.~\ref{sec:conclusions}.

\section{Theoretical description}\label{sec:theoreticalPredictions}

\begin{figure}
	\begin{center}
		\includegraphics[scale=1.4]{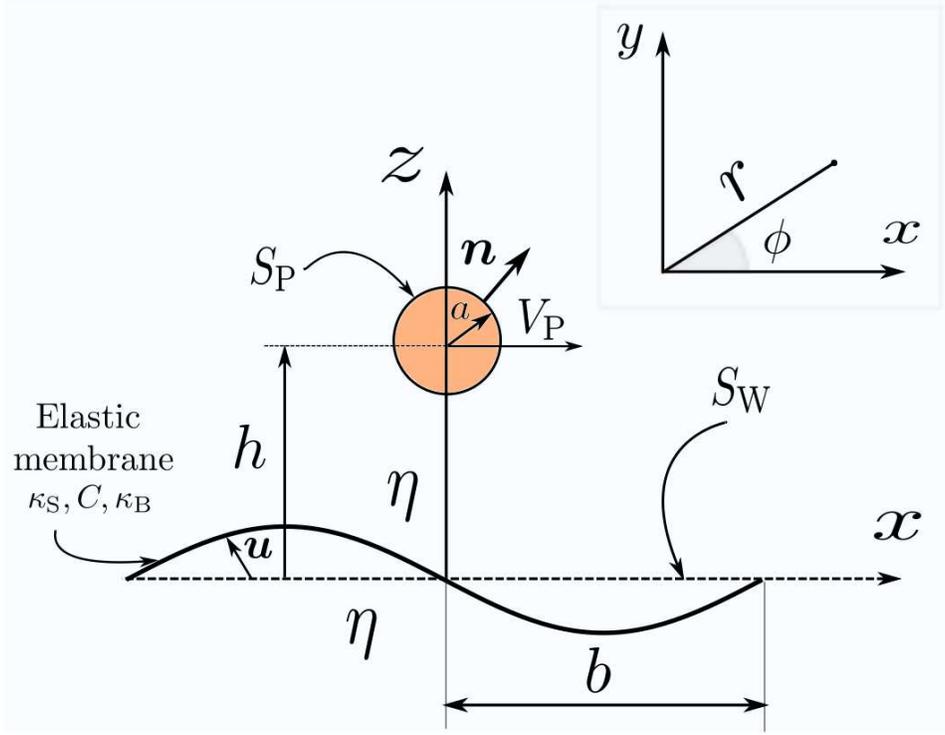}
		\caption{Illustration of the elastohydrodynamic problem. 
		A solid sphere of radius~$a$ located a distance~$h$ above an elastic membrane of radius~$b$.
		In the undeformed state, the membrane is extended in the plane $z=0$.
		The frame of reference attached to the center of the sphere translates at a constant velocity~$\vect{V}_\mathrm{P}$ with respect to the laboratory frame. 
		The fluid on both sides of the membrane has the same dynamic viscosity~$\eta$.
		The figure in the inset is a top view of the frame of reference associated with the particle where $(r,\phi)$ are the polar coordinates. }
		\label{illustration}
	\end{center}
\end{figure}

We consider the quasi-steady motion of a solid spherical particle of radius~$a$, initially located at position $z=h$ above a finite-sized elastic membrane of radius $b$ extended in the $xy$ plane; the $z$ direction is normal to the plane.
The particle translates at a constant velocity $\vect{V}_\mathrm{P} = V_\mathrm{P} \vect{e}_x$ parallel to the membrane,  as measured in the laboratory reference frame, schematically illustrated in Fig.~\ref{illustration}.
In the following, we examine the system behavior in the far-field limit such that $a\ll h$.
The fluid on both sides of the membrane is assumed to be Newtonian and the flow is incompressible, characterized by a constant dynamic viscosity~$\eta$.
The membrane is modeled as a two-dimensional sheet made by a hyperelastic material that exhibits resistance toward shear and bending.
Membrane shear elasticity is described by the well-established Skalak model~\cite{skalak73}, which is often used as a practical model for red blood cell membranes~\cite{foessel11, dupont15, barthes16}.
The Skalak model is characterized by the shear modulus~$\kS$ and the area dilatation modulus~$\kA$, which are related by the coefficient $C:=\kA/\kS$. 
The membrane resistance toward bending is described by the Helfrich model~\cite{helfrich73, guckenberger16, guckenberger17}, with the corresponding bending modulus~$\kB$.
For small membrane displacements away from a plane, the linearized traction jump equations stemming from these two models are given by \cite{daddi16, daddi16b}
\begin{subequations}\label{boundaryConditions}
	\begin{align}
	-\frac{\kappa_\mathrm{S}}{3} \big( \Delta_{\parallel} u_\beta + (1+2C) \epsilon_{,\beta} \big) &= \bigtriangleup    f_{\beta} \, , \quad \beta \in \{x,y\} \, , \label{sigma_tangential}  \\
	 \kappa_\mathrm{B}   \Delta_{\parallel}^2 u_z &= \bigtriangleup    f_{z}  \quad \text{on} \quad \xS \, , \label{sigma_normal}
	\end{align}
\end{subequations}
where $\vect{u}$ is the displacement vector of the material points of the membrane relative to their initial positions, and $\xS = x\vect{e}_x + y \vect{e}_y$ denotes the position vector of the material points relative to the \emph{planar} configuration of reference.
Here $\Delta_\parallel$ denotes the Laplace-Beltrami operator~\cite{deserno15}, defined for a given scalar function~$w$ as $\Delta_\parallel w := w_{,xx} + w_{,yy}$, $\epsilon:= u_{x,x}+u_{y,y}$ is the dilatation, and $\bigtriangleup   \vect{f}$ stands for the traction jump across the membrane.
Note that a comma in indices means a partial spatial derivative.

It is convenient to describe the present problem in a translating reference frame attached to the sphere, in which the fluid far away from the sphere translates with velocity $-\bm{V}_P$
The fluid velocity and stress fields, $\bm{v}(\bm{x})$ and $\bm{\sigma}(\bm{x})$, respectively, satisfy the continuity and Stokes equations~\cite{happel12}
\begin{equation}
 \nabla \cdot \bm{v} = 0 \, ,  \qmbox{and} \nabla \cdot \bm{\sigma} = \bm{0} \, ,
\end{equation}
and the boundary conditions 
\begin{equation}
 \bm{v}|_{S_\mathrm{P}} = \bm{0} \, , 
  \qmbox{and} \bm{v}|_{S_{\infty}} = - \bm{V}_\mathrm{P} \, , 
\end{equation}
where $S_{\infty}$ is a bounding surface at infinity, and $S_\mathrm{P}$ denotes the surface of the particle.
Moreover, $\sig = -p \vect{I} + 2\eta \vect{E}$ is the fluid stress tensor with $p$ denoting the pressure and $\vect{E} = \tfrac{1}{2} \left( \bNabla \vv + \bNabla \vv^{\mathrm{T}} \right)$ is the rate-of-strain tensor.
The traction jumps appearing on the right hand side of Eqs.~\eqref{boundaryConditions} are related to the stress tensor via the relation $\bigtriangleup f_\beta = \sigma_{z\beta} (z=0^+) - \sigma_{z\beta} (z=0^-)$, for $\beta \in \{x,y,z\}$.

The no-slip boundary condition at the deformed membrane provides a direct link between the membrane displacement~$\bm{u}$ and fluid velocity~$\bm{v}$.
Specifically, 
 \begin{equation}
 \left. \frac{\operatorname{D} \uu}{\operatorname{D}t} 
:= \frac{\partial \uu}{\partial t} + \vv \cdot \bNablaParallel \uu 
 = (\vv + \bm{V}_\mathrm{P}) \right|_{\xS + \uu (\xS)} \, , \label{materialDerivativeOfU}
\end{equation} 
where $\bNablaParallel = \vect{e}_x \partial_{x} + \vect{e}_y \partial_{y}$ is the tangential gradient operator taken along the membrane.
In this paper, we consider a small but finite deformation amplitude relative to the distance between the particle and the membrane ($|\bm{u}| \ll h$). Then, the no-slip condition \eqref{materialDerivativeOfU} can be mapped onto the reference plane $\bm{x}_\mathrm{S}$ by using a Taylor expansion to write $\bm{v}|_{\bm{x}_\mathrm{S} + \bm{u}(\bm{x}_\mathrm{S})} = \bm{v}|_{\bm{x}_\mathrm{S}} + \bm{u} \cdot \nabla \bm{v} + O(\bm{u} \bm{u} : \nabla \nabla \bm{v})$. In the limit of a quasisteady membrane displacement ($\pp{\bm{u}}{t} = 0$) as measured in the translating reference frame,  substituting the above expansion into \eqref{materialDerivativeOfU} gives
\begin{equation}
\vv = -\bm{V}_\mathrm{P} + \bm{V}_\mathrm{W} \quad \text{on} \quad \xS \, \qmbox{where}  \bm{V}_\mathrm{W} =    \vv \cdot \bNablaParallel \uu - \uu \cdot \bNabla \vv \, , \label{velocityFromTaylorexpansion}
\end{equation}
which an effective boundary condition prescribed on a planar (undeformed) wall.

\section{Reciprocal theorem}\label{sec:reciprocalTheorem} 

Before attempting to solve the problem at hand, of a sphere translating tangent to an elastic membrane, we introduce a related problem, namely that of a particle of surface $S_\mathrm{P}$ translating with velocity $\bm{V}_\mathrm{P}$ tangent to a planar wall~$S_\mathrm{W}$ with a prescribed surface velocity distribution $\bm{V}_{\mathrm{W}}(\bm{x}_\mathrm{S})$, with $\bm{x}_\mathrm{S} \in S_\mathrm{W}$. Note that both $\bm{V}_\mathrm{P}$ and $\bm{V}_\mathrm{W}$ may be arbitrarily oriented relative the the surface $S_{\mathrm{W}}$. We are interested in the relationship between the hydrodynamic force acting on such a particle, its translational velocity $\vect{V}_\mathrm{P}$, and the prescribed surface velocity $\bm{V}_{\mathrm{W}}(\bm{x}_\mathrm{S})$.

An explicit expression for the force on the particle may be obtained using the Lorentz reciprocal theorem for Stokes flows. To this end, we define a model problem wherein a particle translates at velocity $\vect{\hat{V}}_\mathrm{P}$ relative to a \emph{rigid} no-slip wall in fluid that is quiescent far away from the sphere (as measured in the laboratory reference frame). In the frame of reference of the sphere, the flow in the model problem is described by a velocity field $\hat{\vect{v}}$ and a stress field $\vect{\hat{\sigma}}$ that satisfy the Stokes equations and similar boundary conditions as above, but involving hatted quantities and with $\bm{V}_\mathrm{W}(\bm{x})$ being absent. Specifically, $\vect{\hat{v}}(\vect{x} \in S_\mathrm{P})=\boldsymbol{0}$, $\vect{\hat{v}}(\vect{x} \in S_\mathrm{W})=-\vect{\hat{V}}_\mathrm{P} $, and $\vect{\hat{v}}(\vect{x} \to \infty)=-\vect{\hat{V}}_\mathrm{P} $.

The Lorentz reciprocal theorem for Stokes flows \cite{happel12} states that 
\begin{equation}
 \int_{S_\mathrm{P} + S_\mathrm{W} + S_{\infty}} \bm{n} \cdot \bm{\sigma} \cdot \hat{\vect{v}}\, \rmd S = \int_{S_\mathrm{P} + S_\mathrm{W} + S_{\infty}} \bm{n} \cdot \vect{\hat{\sigma}} \cdot \bm{v}\, \rmd S \, .
\end{equation}
On applying the boundary conditions and using the definition for the hydrodynamic force, 
\begin{equation}
\bm{F}^\mathrm{H} = \int_{S_\mathrm{P}} \bm{n} \cdot \bm{\sigma} \, \rmd S = -\int_{S_\mathrm{W} + S_{\infty}} \bm{n} \cdot \bm{\sigma} \, \rmd S \, , 
\end{equation}
we obtain
\begin{equation}
 -\vect{\hat{F}}^\mathrm{H} \cdot \bm{V}_\mathrm{P} + \bm{F}^\mathrm{H} \cdot \vect{\hat{V}}_\mathrm{P} = \int_{S_\mathrm{W}} \bm{n} \cdot \vect{\hat{\sigma}} \cdot \bm{V}_\mathrm{W}\, \rmd S \, , \label{lotenzDemiFinal}
\end{equation}
where $\vect{\hat{F}}^\mathrm{H}$ is the hydrodynamic force in the model problem. The above expression lets us compute the projection of the hydrodynamic force on the particle in the direction of the arbitrarily chosen vector $\vect{\hat{V}}_\mathrm{P}$ for a specified $\bm{V}_\mathrm{W}(\bm{x}_\mathrm{S})$, assuming that the stress field in the model problem is fully known. 

We now specialize the general expression above to the case of a spherical particle translating parallel to a planar wall, $\bm{V}_\mathrm{P} = V_\mathrm{P} \bm{e}_x$. Here, the wall coincides with the $xy$ plane and the $z$ axis points towards the particle center, as shown in Fig.~\ref{illustration}. 
In particular, we are interested in the wall-normal component of the force acting on the sphere $(F_{\perp}^\mathrm{H} = \bm{F}^\mathrm{H} \cdot \ez)$.
For $\vect{\hat{V}}_\mathrm{P} = \hat{V}_\mathrm{P} \ez$, the first term in~\eqref{lotenzDemiFinal} drops out (here $\vect{n} = \vect{e}_z $ since $S_\mathrm{W}$ is a plane), leading to
\begin{equation}
 F_{\perp}^\mathrm{H} = \frac{1}{\hat{V}_\mathrm{P}} \int_{S_\mathrm{W}} \bm{n} \cdot \vect{\hat{\sigma}} \cdot \bm{V}_\mathrm{W}\, \rmd S = \frac{1}{\hat{V}_\mathrm{P}}\int_{S_\mathrm{W}} \left(\hat{\sigma}_{zz} V_{Wz} + \hat{\sigma}_{zx} V_{Wx} + \hat{\sigma}_{zy} V_{Wy} \right)\, \rmd S \, .
 \label{lorenzFinal}
\end{equation}
Whether or not the components of the surface velocity $\bm{V}_\mathrm{W}$ contribute to the normal force can be established by their spatial symmetry relative to components of $\ez \cdot \bm{\sigma}$. For example, $\hat{\sigma}_{zz}$ is an even function of both $x$ and $y$ due to the axisymmetry of the model problem, therefore normal velocity distributions $V_{Wz}(x,y)$ that share this symmetry can contribute to $F_{\perp}$. Similarly, $V_{Wx}(x,y)$ distributions that are odd in~$x$ and even in~$y$, and $V_{Wy}(x,y)$ distributions that are odd in $y$ and even $x$, can contribute to a normal force. As we will show below, all three symmetries are realized when a sphere translates parallel to a membrane that resists stretching and bending. While it has been reported in other contexts~\cite{sekimoto93, skotheim04, skotheim05} that out-of-plane deformation (here mediated by bending) can produce a normal force, we will show that an in-plane stretching has an opposite effect. We will quantify these findings in subsequent sections.

\section{Calculation of the normal force}\label{sec:calculations} 

\subsection{Rescaling}

Having derived a general reciprocal relation for the normal force on a particle translating tangent to a surface with a prescribed surface velocity $V_W(\bm{x}_S)$, we now compute specific results for the force when this surface velocity is a result of the elasticity of the membrane.
For that purpose, it is convenient to rescale the system properties in the main and model problems by introducing dimensionless variables, which we denote by a star.

We observe from \eqref{velocityFromTaylorexpansion} that the velocity scale at the membrane is $V_{\rm P}$. However, this scale corresponds to uniform translation (due the choice of reference frame), and is therefore not associated with velocity gradients or fluid stresses. The fluid stress is a result of the disturbance flow due to the translating particle, which, near the membrane, has a characteristic velocity $V_{\rm P} a/h$ that decays over a characteristic length scale $h$. The stress acting on the membrane therefore has the characteristic scale $\eta V_{\rm P} a/h^2$. Analogous relations apply for the model problem. Accordingly, we define dimensionless variables
\begin{equation}
 \vv = \frac{a \Vpcom}{h} \, \vv^\star , \quad \sig = \frac{a\eta \Vpcom}{h^2} \, \sig^\star \, , \quad
 \hat{\vv} = \frac{a\hat{V}_\mathrm{P}}{h} \, \hat{\vv}^\star , \quad \hat{\sig} = \frac{a\eta \hat{V}_\mathrm{P} }{h^2} \, \hat{\sig}^\star \, , \label{rescaling}
\end{equation} 
and  rescale all lengths of the problem by~$h$.  
In this paper, we focus our attention on the far-field limit where $a/h \ll 1$.
By examining the boundary conditions prescribed at the membrane in~\eqref{boundaryConditions}, it can be noted that the linearized tangential traction jumps at the membrane are imposed by shear resistance only and involve second-order derivatives of the in-plane displacements.
In contrast, the linearized normal traction jump is imposed by bending resistance only and involves fourth order derivatives of the out-of-plane displacement.
Based on these considerations and using the stress scale $\eta V_{\rm P} a/h^2$, we define the rescaled membrane displacements as follows
\begin{equation}
 u_x = \frac{a \eta \Vpcom}{\kS} \, u_x^\star \, , \quad 
 u_y = \frac{a\eta \Vpcom}{\kS} \, u_y^\star \, , \quad 
 u_z = \frac{a\eta \Vpcom h^2}{\kB} \, u_z^\star \, . \label{ux_uy_uz}
\end{equation}
 
In the limit of small membrane deformation $(\bm{u} \ll h)$, the present elastohydrodynamic problem can conveniently be solved perturbatively.
We define the perturbation parameters 
\begin{equation}
 \LambdaS = \frac{a\eta \Vpcom}{h\kS} \, , \qquad
 \LambdaB = \frac{a\eta \Vpcom h}{\kB} \, , \label{definitionLambda}
\end{equation}
which can be regarded as dimensionless compliances associated with the membrane resistance toward shear and bending, respectively.
Using \eqref{ux_uy_uz} and \eqref{definitionLambda}, we can write the membrane displacement vector as
\begin{equation}
 \uu =  h \LambdaS \left( u_x^\star \vect{e}_x + u_y^\star \vect{e}_y \right) + h \LambdaB u_z^\star \vect{e}_z \, .
\end{equation} 
Note that $\LambdaB  = 0$ for an idealized membrane with pure shear (such as that of an artificial capsule designed for drug delivery) and $\LambdaS = 0$ for a membrane with pure bending (such as that of a fluid vesicle or a liposome).
For a particle-membrane distance $h=\left(\kB/\kS\right)^{1/2}$ both dimensionless numbers $\LambdaS$ and $\LambdaB$ are equal.
This corresponds to the situation where shear and bending equally manifest themselves in the system~\cite{daddi18epje}.

\subsection{Perturbation solution}

In order to obtain approximate analytical expressions for the induced normal force $F_\perp^\mathrm{H}$ acting on the translating particle, we will focus our attention to the limit of small membrane deformation, so that $\LambdaS \ll 1$ and $\LambdaB \ll 1$.
We can thus expand perturbatively the velocity and displacement fields in power series of the dimensionless numbers~$\LambdaS$ and $\LambdaB$.
To leading order, the rescaled displacement and velocity fields can be written using a regular perturbation expansion as
\begin{equation}
	\uu^\star = \uu_0^\star + O(\LambdaB,\,\LambdaS) \, , \qquad
		 \vv^\star = \vv_0^\star + O(\LambdaB,\,\LambdaS)\, , \label{UAndV_Perturb}
\end{equation}
where $\uu_0^\star$ and $\vv_0^\star$ are the solutions of the zeroth-order problem corresponding to a planar undeformed membrane. 
From the boundary condition~\eqref{velocityFromTaylorexpansion} imposed at the undisplaced (planar) membrane, it follows readily that $\bm{v}_0^\star = - \bm{V}_\mathrm{P}^\star = -(h/a) \bm{e}_x$ on the planar surface of reference~$\X_\mathrm{S}$.
Substituting Eqs.~\eqref{UAndV_Perturb} into~\eqref{lorenzFinal} and keeping only the leading order terms in~$\LambdaB$ and $\LambdaS$, the hydrodynamic force exerted on the particle translating parallel to the membrane simplifies to 
\begin{align}
 F_{\perp}^\mathrm{H} &= -\eta a V_\mathrm{P} 
 \int_{S_\mathrm{W}} \left[ \LambdaB \left\{
  \frac{\partial {u_z}_0^\star}{\partial x^\star} \, \hat{\sigma}_{zz}^\star + a^\star {u_0^\star}_z \left(\frac{\partial {v_x}_0^\star}{\partial z^\star} \hat{\sigma}_{zx}^\star + \frac{\partial {v_y}_0^\star}{\partial z^\star} \hat{\sigma}_{zy}^\star\right)\right\}  + \LambdaS \left\{\frac{\partial {u_x}_0^\star}{\partial x^\star}  \hat{\sigma}_{zx}^\star
 +  \frac{\partial {u_y}_0^\star}{\partial x^\star}  \hat{\sigma}_{zy}^\star \right\} \right] \rmd S^\star, 
 \label{firstOrderProblemVanishingFreq_simplifed}
\end{align}
where $a^\star = a/h$.
This is a central result of our paper that we evaluate below.
It is worth noting that on $\xS$, both of the partial derivatives $\partial \vect{v}_0^\star / \partial x^\star $ and $\partial \vect{v}_0^\star / \partial y^\star $ vanish.
Due to the decoupled nature between the shear and bending deformation modes, the solution of the flow problem near a membrane endowed simultaneously with both shear and bending resistances can readily be obtained via linear superposition of the two independent  shear and bending contributions.

Consequently, the normal force is found to scale quadratically with the particle velocity on account of the fact that $\LambdaB$ and $\LambdaS$ are linear in $V_\mathrm{P}$.
This situation is in contrast to that of the drag force which is known to scale linearly with velocity.
Notably, the normal force equals to zero near an undeformed, planar wall, which corresponds to an elastic membrane with infinite shear and bending moduli where $\LambdaB\to 0$ and $\LambdaS\to 0$.

For $a \ll h$, the fluid stress tensor in the model problem of a sphere moving perpendicular to a no-slip wall can be obtained to leading order in particle radius using the method of images due to Blake~\cite{blake71}.
For an infinitely-extended rigid wall, the normal components of the stress tensor in the cylindrical coordinate system, 
are given by~\cite{lee79}
\begin{equation}
	\hat{\sigma}_{zz}^\star = \frac{9 \left(1+\frac{9}{8} \, a^\star\right)}{\left(1+r^2\right)^{5/2}} \, , \qquad
	\hat{\sigma}_{zr}^\star = -\frac{9\left(1+\frac{9}{8} \, a^\star\right) r}{\left(1+r^2\right)^{5/2}}  \,
	\quad \text{on}\quad \xS \, , 
	\label{stressTensorHardWall}
\end{equation}
wherein $r$ is the radial distance measured in the comoving frame of reference translating at the particle velocity (c.f.~inset of Fig.~\ref{illustration}).
We further note that  $\hat{\sigma}_{zx}^\star = \hat{\sigma}_{zr}^\star \cos\phi$ and $\hat{\sigma}_{zy}^\star = \hat{\sigma}_{zr}^\star \sin\phi$ where $\phi \in [0,2\pi]$ is the polar angle.

We now assume that the particle is located at the center of a membrane of dimensionless radius~$b^\star$.
Even though \eqref{stressTensorHardWall} applies in principle to an infinitely-extended rigid wall $(b^\star\to\infty)$, we will assume in the sequel that these expressions approximately hold for a finite-sized disk provided that $b^\star \gg 1$.
It can be noticed that in the far-field limit, the $zz$ and $zr$ components of the fluid stress tensor stated above undergo a rapid decay with distance as $r^{-5}$ and~$r^{-4}$, respectively.
Consequently, for relatively large membrane sizes our simplifying approximation should be reasonable.
An exact analytical solution of the axisymmetric flow problem due to a Stokeslet directed along the axis of a circular hard disk has been previously obtained in the form of a dual integral equation~\cite{kim83jpsj}, finding that the wall-induced correction to the hydrodynamic drag force exerted on a sedimenting particle decays with disk radius as ${b^\star}^{-5}$ and approaches rapidly the result by Lorentz~\cite{lorentz07, lee79} for moderately large values of $b^\star$. 
Throughout this manuscript, we will thus assume that the membrane size is sufficiently large for the above approximation to be valid.

\subsection{Bending- and shear-related contributions to the lift force}

We will consider next the bending- and shear-related contributions to the normal force separately. 
As previously mentioned, the membrane normal displacement $u_z^\star$ is a function of the membrane bending properties only and does not depend on shear. 
It is straightforward, though tedious, to calculate the solution in the zeroth-order problem for the normal displacement.
As derived in the Appendix, the normal displacement for a finite-sized membrane can be presented in the form
\begin{equation}
 \begin{split}
 {u_0}_z^\star &= H(r) \cos\phi \, , \label{u_z_bending}
 \end{split}
\end{equation}
where $H$ is a radial function that satisfies the boundary conditions of vanishing displacement and slope at $r=b^\star$, given explicitly by Eq.~\eqref{H}. 
The derivative of the normal displacement with respect to $x^\star$ which is required for the application of the reciprocal theorem is explicitly given by Eq.~\eqref{uz_x_Appendix}.

In addition, the solution of the zeroth-order problem for the in-plane displacements due to shear for a finite-sized membrane can be cast in the form
\begin{equation}
	{u_0}_x^\star (r,\phi) = A(r) \cos (2 \phi) + G(r) \, , \qquad
		 {u_0}_y^\star (r,\phi) = A(r) \sin (2\phi) \, , 
\end{equation}
where the radial functions $A$ and $G$ are given by Eqs.~\eqref{AG} and satisfy $A(r=b^\star) = G(r=b^\star) = 0$ to ensure zero displacement at the membranes extremities.
The derivatives of the in-plane displacements with respect to~$x^\star$ are given by Eqs.~\eqref{ux_x_uy_x_Appendix} of the Appendix. Finally, the radial velocity gradient at the elastic membrane in the zeroth-order problem can readily be determined from the solution for the flow field near a planar undeformed wall and is found to be~\cite{lee79}
\begin{align}
	\left. \frac{\partial {v_0^\star}_r}{\partial z^\star} \right|_{z=0} &=  \frac{9 \left(1+\frac{9}{16} \, a^\star\right) r^2}{\left(1+r^2\right)^{5/2}}  \cos \phi \, . \label{veloGrad_r} 
\end{align}

Next, we substitute Eqs.~\eqref{stressTensorHardWall} through \eqref{veloGrad_r} into the integral equation giving the hydrodynamic normal force~\eqref{firstOrderProblemVanishingFreq_simplifed}.
Passing to polar coordinates yields the expressions of the bending- and shear-related contributions to the normal force.
Specifically, 
\begin{subequations}
	\begin{align}
		{F}_{\perp, \mathrm{B}}^\mathrm{H}  &= 
		-\eta a V_\mathrm{P} \, \LambdaB
		\int_0^{2\pi} \int_0^{b^\star} 
		\left( \hat{\sigma}_{zz}^\star \frac{\partial {u_0}_z^\star}{\partial x^\star} + a^\star \hat{\sigma}_{zr}^\star {u_0}_z^\star \, \frac{\partial {v_0}_r^\star}{\partial z^\star}
		 \right) \, r\,  \Intd r \, \Intd \phi \, , \label{Normal-force-Bending} \\
		{F}_{\perp, \mathrm{S}}^\mathrm{H}  &= 
			-\eta a V_\mathrm{P}\, \LambdaS
			 \int_0^{2\pi}  \int_0^{b^\star}
			\left( \hat{\sigma}_{zx}^\star \frac{\partial {u_0}_x^\star}{\partial x^\star} 
			+ \hat{\sigma}_{zy}^\star \frac{\partial {u_0}_y^\star}{\partial x^\star} 
			 \right)  \, r \, \Intd r \, \Intd \phi \, , \label{Normal-force-Shear} 
	\end{align}
\end{subequations}
which, upon integration leads to the final analytical expressions evaluated up to terms of $\mathcal{O}({a^\star}^2)$, 
\begin{subequations}\label{lorentzRecThm_scalar}
	\begin{align}
	 {F}_{\perp, \mathrm{B}}^\mathrm{H}   &= \eta a V_\mathrm{P} \, \LambdaB 
	 \Big( \left(1+\tfrac{27}{16} \, a^\star\right)I_1 - a^\star I_2 \Big) \, , \label{liftForce-bending-finiteSize} \\
	 {F}_{\perp, \mathrm{S}}^\mathrm{H}   &=  -\eta a V_\mathrm{P} \, \LambdaS \left(1+\tfrac{27}{16} \, a^\star\right) I_3 \, .
	\end{align} 
\end{subequations}
Here, the quantities $I_\alpha>0, \, \alpha\in \{1,2,3\}$ depend on membrane size and can conveniently be expressed as functions of the parameter $\lambda := \left( 1+{b^\star}^2 \right)^{1/2}$ as
\begin{subequations}\label{I-expressions}
	\begin{align}
	 I_1 &= \frac{9\pi}{2} \Bigg( 2\ln \left( \frac{(1+\lambda)^2}{4\lambda} \right) 
	  -\frac{(\lambda^2+2\lambda-1)(\lambda-1)^2}{\lambda^2(\lambda+1)^2} \Bigg) \, , \\
	 I_2 &= \frac{27\pi}{320} \bigg( 60\ln \lambda -113+\frac{180}{\lambda} 
	   - \frac{60}{\lambda^2}-\frac{40}{\lambda^3} + \frac{45}{\lambda^4} - \frac{12}{\lambda^5} \bigg) \, , \\
	 I_3 &= \frac{27\pi}{4 (1+C)} \frac{(2\lambda+1)(\lambda-1)^3}{\lambda^6 (\lambda+1)^2} 
	      \left( 1 + 3\lambda + \frac{4(1+2C)\lambda^2}{3+2C} \right) \, ,
	\end{align} 
\end{subequations}
where $C=\kA/\kS$ and appears in the shear contribution to tangential stress balance at the membrane.  
Note that $I_1$ and $I_2$ are associated with the contributions originating from the first and second integrals of Eq.~\eqref{Normal-force-Bending}, respectively.
While the first term leads to a positive contribution to the lift force, the second term is found to have an opposite effect.
However, since $I_1 > I_2$ for all values of $\lambda \ge 1$, and that $a^\star \ll 1$, the resulting normal lift force is always directed away from the membrane.
In contrast, the shear-related contribution to the lift force has an opposite effect, leading to an attraction of the particle toward the membrane.
A similar behavior has previously been observed when two particle are set into motion toward an elastic membrane, where  bending rigidity always leads to mutual repulsion whereas shear resistance can lead to attractive interaction~\cite{daddi16c}.

For a very large membrane, say $\lambda \gg 1$, the rescaled lift forces due to bending and shear have the asymptotic form 
\begin{subequations}
	\begin{align}
		\frac{{F}_{\perp, \mathrm{B}}^\mathrm{H}}{\eta a V_\mathrm{P} \, \LambdaB } &= 
	     9\pi \left( (1+ \tfrac{9}{8}\, a^\star)\ln\lambda
	    -2(1+\tfrac{27}{16}\, a^\star) \ln 2 - \tfrac{1}{2} + \tfrac{69}{320}\, a^\star   \right)  
	     \, , \label{asymptotisch} \\
		\frac{{F}_{\perp, \mathrm{S}}^\mathrm{H}}{\eta a V_\mathrm{P} \LambdaS}  &=   -\frac{54\pi(1+2C) \left(1+\tfrac{27}{16} \, a^\star\right)}{(3+2C)(1+C)\lambda^2} + \mathcal{O} \left(\lambda^{-3}\right) \, .		
	\end{align}
\end{subequations}

It can clearly be seen that the bending-related contribution to the normal force diverges logarithmically with the membrane size, whereas the shear-induced normal force decays as $\lambda^{-2}$ and eventually vanishes as the membrane radius goes to infinity. 
Notably, as $C\to\infty$, which corresponds physically to an incompressible membrane, the shear-induced normal force vanishes for all membrane sizes.

\begin{figure}
\begin{center}
\scalebox{0.9}{\input{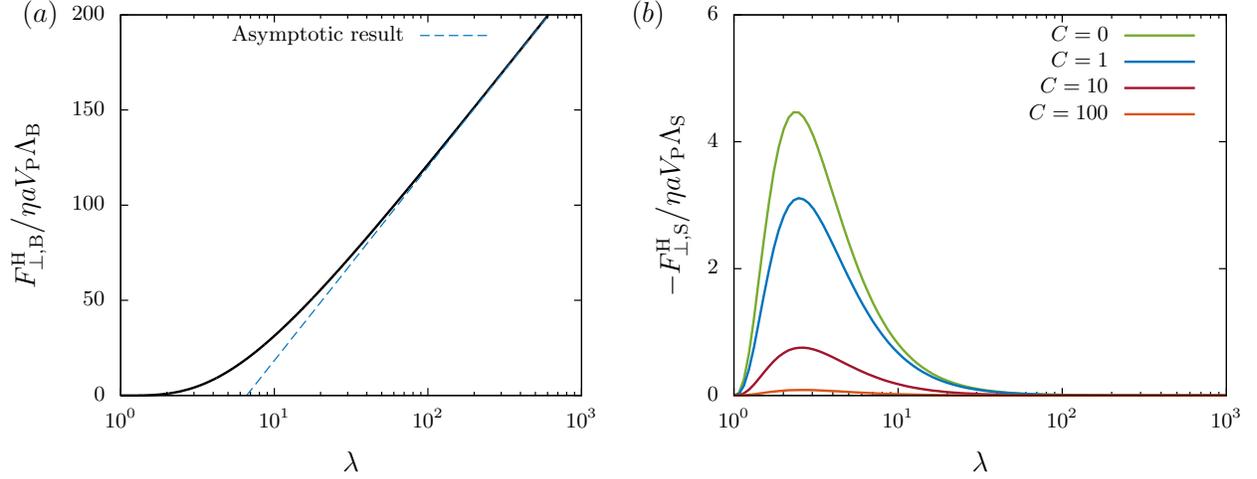}}
\end{center}
\caption{(Color online) Variation of the rescaled normal lift force due to particle motion tangent to a finite-sized membrane with $(a)$ pure bending and $(b)$ pure shear versus the parameter $\lambda= \left( 1+b^2 \right)^{1/2}$ as predicted theoretically from Eqs.~\eqref{lorentzRecThm_scalar}--\eqref{I-expressions}.
The blue dashed line shown in $(a)$ is the asymptotic results given by Eq.~\eqref{asymptotisch}.
}
\label{Lift_BenShe}
\end{figure}

In Fig.~\ref{Lift_BenShe} we illustrate the variation of the dimensionless lift force induced on a solid particle translating parallel to an idealized elastic membrane with $(a)$ pure bending and $(b)$ pure shear as a function of the system size parameter~$\lambda$.
Results for four values of the Skalak ratio~$C$ are shown which span the most likely values for elastic membranes to be expected for a wide range of situations.
Qualitatively, in the range of the present analytical theory $(\lambda \gg 1)$, lower values of $\lambda$ correspond to small normal forces, and vice versa.

In typical blood flow situations~\cite{Freund_2014}, red blood cells have a radius of $b=5 \times 10^{-6}$m, bending modulus $\kB=2\times 10^{-19}$Nm, shear modulus $\kS = 5\times 10^{-6}$N/m with $C=100$.
According to our analytical predictions, a spherical particle of radius $a=0.15\times 10^{-6}$m, which is located above a cell membrane at a distance $h=0.3\times 10^{-6}$m (leading to a dimensionless membrane radius of $b^\star \simeq 17$), translating at velocity $V_\mathrm{P}=10^{-6}$m/s in a fluid of dynamic viscosity $\eta=1.2 \times 10^{-3}\operatorname{Pa}\cdot\operatorname{s}$ will experience a lift force of about $0.1~\%$ of the opposing drag force.
This induced lift force is bending dominated as the effect of shear dies out rapidly for a large system size.

Since the bending-related contribution to the normal force diverges logarithmically as $\lambda\to\infty$, we will present in the following section a regularization procedure to yield a finite lift force near an infinitely-extended membrane.
A similar regularization approach has previously been employed by Bickel to investigate the Brownian motion near a liquid-like membrane~\cite{bickel06}, or the hydrodynamic mobility near a deformable fluid interface~\cite{bickel07}.

\section{Regularization solution }\label{sec:regularization} 

We regularize the bending operator by introducing a length scale $\epsilon^{-1}$ beyond which bending becomes subdominant to a body force~\cite{bickel07}, e.g.\@ gravity, such that $\epsilon^{-4} = \kB/g\Delta\rho$, where $g$ is the acceleration due to gravity and $\Delta\rho$ is the density difference between the lower and the upper phases. 
Accordingly, the rescaled membrane normal displacement in the zeroth-order problem is the solution of  the regularized biharmonic equation (c.f.\@ Bickel, Ref.~\onlinecite{bickel06})
\begin{equation}
	\left( \Delta_\parallel^2 + \epsilon^4 \right) {u_0}_z^\star = {\sigma_0}_{zz}^\star \, , \label{bending-regularized}
\end{equation}
wherein
\begin{equation}
	{\sigma_0}_{zz}^\star = -\frac{9\left(1+\frac{9}{16} \, a^\star\right) r}{\left(1+r^2\right)^{5/2}}  \cos\phi \, .
\end{equation}
Here, we restrict our attention for simplicity to particle motion tangent to an infinitely-extended membrane for which the normal force is shown in the previous section to be logarithmically divergent.
It is more convenient to solve the above equation using a Fourier transform technique and employing Parseval's theorem.
We define the 2D Fourier transform~\cite{bracewell99}
\begin{equation}
\mathscr{F} \{ f(\vect{x}) \} =: \tilde{f} (\vect{q}) =
\int_{\mathbb{R}^2} f(\vect{x}) e^{-i \vect{q} \cdot \vect{x}} \, \Intd \vect{x} \, , 
\label{2DFourierTransform}
\end{equation}
where  $\vect{x} = (x,y)$ is the projection of the position vector $\vect{r}$ onto the horizontal plane, and $\vect{q} = (q\cos\theta, q\sin\theta)$ is the wavevector that sets the coordinates in Fourier space.
In addition, we recall Parseval's theorem which relates the product of two functions in the real domain to that in the wavenumber domain,~\cite{chew95}
\begin{equation}
	\int_{\mathbb{R}} f(\vect{x}) \, g(\vect{x}) \, \Intd \vect{x}  
	= \frac{1}{(2\pi)^2} \int_{\mathbb{R}} \tilde{f}(\vect{q}) \, \left\{\tilde{g}(\vect{q})\right\}^* \, \Intd \vect{q} \, ,
	\label{Parseval}
\end{equation}
where an asterisk denotes a complex conjugate.
Applying the identity~\eqref{Parseval} to Eq.~\eqref{Normal-force-Bending}, which provides the bending-related contribution to the normal lift forces yields
\begin{equation}
	{F}_{\perp, \mathrm{B}}^\mathrm{H}  =  -\frac{\eta a V_\mathrm{P}}{(2\pi)^2} \, \LambdaB \int_0^{2\pi}  \int_0^\infty 
	 \left( iq\cos\theta \, \left\{\widetilde{\hat{\sigma}_{zz}^\star} \right\}^* 
	 + 
	 a^\star \left\{\widetilde{ \hat{\sigma}_{zr}^\star \frac{\partial {v_0^\star}_r}{\partial z^\star}} \right\}^* \right)
	 \widetilde{{u_0}_z^\star} \, q\, \Intd q  \, \Intd \theta  \, .
	 \label{Bending_Parseval}
\end{equation}

Next, by transforming Eq.~\eqref{bending-regularized} into Fourier space, and making use of the equality
\begin{equation}
\int_0^{2\pi} \cos\phi \, e^{-iqr\cos(\phi-\theta)} \, \Intd \phi 
= -2i\pi \cos\theta J_1(qr) \, ,
\end{equation}
where $J_1$ is the Bessel function of the first kind, the normal displacement of the membrane is expressed in Fourier space by
\begin{equation}
	\widetilde{{u_0}_z^\star} = \frac{\widetilde{{\sigma_0}_{zz}^\star}}{q^4+\epsilon^4} 
	=  \frac{6i\pi \left(1+\tfrac{9}{16}\, a^\star \right) q }{q^4+\epsilon^4} \, e^{-q} \cos\theta \, .
	\label{uZ_FourierSpace}
\end{equation} 
Evidently, at large distances $q \ll \epsilon \ll 1$, the deformation decays to zero, and thus the Fourier transform of~${u_0}_z^\star$ is well defined. 
Since $\hat{\sigma}_{zz}^\star$ is a radially-symmetric function in $r$ (c.f.~\eqref{stressTensorHardWall}), its 2D Fourier transform is simply the zeroth-order Hankel transform apart from a factor $2\pi$, which readily leads to
\begin{equation}
	\widetilde{\hat{\sigma}_{zz}^\star} = 6\pi \left(1+\tfrac{9}{8}\, a^\star \right) (1+q)e^{-q} \, .
\end{equation}
In addition, it follows from Eqs.~\eqref{stressTensorHardWall} and \eqref{veloGrad_r} that
\begin{equation}
	 \hat{\sigma}_{zr}^\star \frac{\partial {v_0^\star}_r}{\partial z^\star} 
	 = -\frac{81 \left(1+\tfrac{27}{16}\, a^\star \right) r^3}{(1+r^2)^5} 
	  \cos\phi \, , \notag 
\end{equation}
the 2D Fourier transform of which is given by
\begin{equation}
	\widetilde{\hat{\sigma}_{zr}^\star \frac{\partial {v_0^\star}_r}{\partial z^\star}}
	= \frac{54i \left(1+\tfrac{27}{16}\, a^\star \right)}{q^4} \, G_q \cos\theta \, , \quad
	\text{where}
	\quad
	G_q := G \left( \left[ \left[\frac{1}{2} \right] , [\,]\right] , 
			\left[ \left[ \frac{9}{2},\frac{5}{2} \right], \left[\frac{3}{2}\right] \right] ,\frac{q^2}{4}  \right) \, .
	\label{sigmaZRtimeVgradient_FourierSpace}
\end{equation}
Here $G$ is the Meijer G-function~\cite{abramowitz72}.
For $q\ll 1$, $G_q \propto q^5$, while for $q \gg 1$, $G_q$ undergoes a rapid exponential decay.
The resulting integral given by \eqref{Bending_Parseval} is thus well behaved and convergent.
By substituting Eqs.~\eqref{uZ_FourierSpace}--\eqref{sigmaZRtimeVgradient_FourierSpace} into~\eqref{Bending_Parseval}, the normal force due to bending, upon regularization for an infinitely-extended membrane, can be presented in a form analogous to \eqref{liftForce-bending-finiteSize} as 
\begin{equation}
	{F}_{\perp, \mathrm{B}}^\mathrm{H}  = \eta a V_\mathrm{P} \, \LambdaB \Big( \left(1+\tfrac{27}{16}\, a^\star \right)I_1' -a^\star I_2' \Big) \, .
	\label{LiftForceRegularization}
\end{equation}
Here $I_1'$ and $I_2'$ are positively defined quantities expressed as integrals over the wavenumber $q$ as
\begin{equation}
	I_1' = 9\pi \int_0^\infty 
				\frac{q^3 (1+q)e^{-2q}}{q^4+\epsilon^4} \, \Intd q \, , \quad\quad
	I_2' = 81 \pi \int_0^\infty  \frac{ G_q e^{-q}}{q^2(q^4+\epsilon^4)} \, \Intd q \, .
\end{equation}
As before, $I_1'$ and $I_2'$ are contributions from the first and second terms in Eq.~\eqref{Bending_Parseval}, respectively, such that $I_1'>I_2'$ for all values of $\epsilon$, and thus leading to a repulsive force.
We further note that both $I_1'$ and $I_2'$ diverge logarithmically as $\epsilon \to 0$.

\begin{figure}
\begin{center}
\scalebox{1.1}{\input{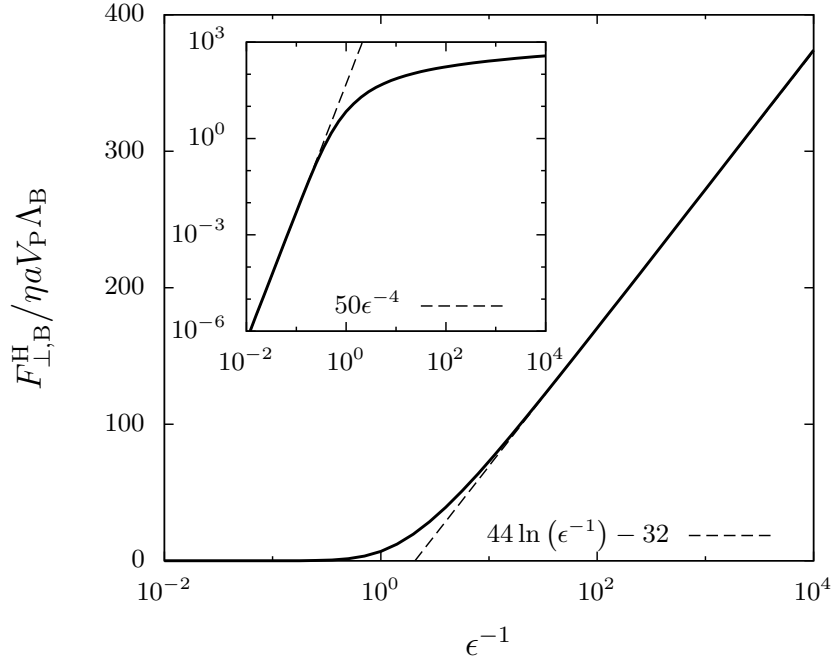}}
\end{center}
\caption{Variation of the rescaled normal lift force due to an infinitely-extended membrane with pure bending versus~$\epsilon^{-1}$ as given by Eq.~\eqref{LiftForceRegularization}.
The dashed line is an asymptotic fit to the lift force in the region $\epsilon^{-1}\gg 1$.
The inset shows the same data in a log-log plot where an asymptotic fit in the region $\epsilon^{-1}\ll 1$ is shown as a dashed line.
}
\label{Lift_Ben_Regularization}
\end{figure}

In Fig.~\ref{Lift_Ben_Regularization} we present the variation of the bending-induced lift force upon regularization versus the cut-off length scale~$\epsilon^{-1}$ stated by Eq.~\eqref{LiftForceRegularization}.
For $\epsilon^{-1} \ll 1$, the body force dominates over the bending force and the normal lift force decays rapidly as~$\epsilon^{-4}$ before it eventually vanishes as $\epsilon^{-1} \to 0$.
In contrast, the lift force increases logarithmically with~$\epsilon^{-1}$ similar to that observed in a finite-sized system shown in Fig.~\ref{Lift_BenShe}$(a)$.
By equating the bending-induced lift force obtained for a finite-sized system
with that calculated in this section using the regularization procedure 
in the limit when $\lambda \gg 1$ and $\epsilon^{-1} \gg 1$, we find that these two systems are in fact equivalents for a cut-off length $ \epsilon^{-1} \simeq b^\star/\pi$.

\section{Conclusions}\label{sec:conclusions} 

In this paper we have derived, using the reciprocal theorem for Stokes flow, expressions for the elastohydrodynamic lift force induced on a spherical particle translating parallel to a realistically-modeled cell membrane possessing resistance toward shear and bending.
Calculations were performed using a far-field model in the point-particle framework valid when the particle radius is small compared to distance from the membrane.
Analytical solutions were derived using a perturbation technique in the small deformation limit.
For a finite-sized membrane of circular shape fixed at its boundaries, the bending- and shear-induced lift forces were determined and expressed in terms of the membrane size in addition to the dimensionless compliances associated with these two deformation modes.
Unlike the viscous drag force, the lift force is found to scale quadratically with particle translational velocity.
Most importantly, the bending-related contribution to the lift force increases logarithmically with the system size whereas shear has an opposite yet insignificant contribution to the total lift force.

A regularization solution was then presented for an infinitely-extended membrane subject to a body force, e.g. gravity, directed along the normal direction.
The lift force was determined analytically using a Fourier transform technique and Parseval's theorem for the resulting integral, and expressed in terms of infinite integrals over the wavenumber.
An analogous logarithmic divergence of the lift force is obtained upon decreasing the cut-off length scale during which the bending forces become dominant over the body force. 
The finite-sized system is found to be asymptotically equivalent to the regularized system in the particular situation where the cut-off length beyond which bending becomes subdominant to a body force is $\epsilon^{-1} \simeq b^\star/\pi$.
Given the far-field approximations made here, there appears to be a small effect on the lift force when considering physical parameters for a typical red blood cell membrane.
For distances very close to the membrane, however, lubrication corrections have to be accounted for, where an enhanced effect is expected.
The membrane-induced lift force quantified in this paper may possibly be of physiological significance for the escape or uptake of targeted viral particles or nanocarriers by the membranes of living cells.

\begin{acknowledgments}
ADMI thanks HAS, BR and the Complex Fluids Group at the Princeton University for their hospitality during a visit when this work was initiated.
HAS and BR are grateful to the NSF for funding via  DMS-1614907.
SG and ADMI thank the Volkswagen Foundation for funding and acknowledge support from the Elite Study Program Biological Physics.
Funding from the DFG (Deutsche Forschungsgemeinschaft) within DA~2107/1-1 (ADMI) is gratefully acknowledged.
This work is supported by the COST Action MP1305, supported by COST (European Cooperation in Science and Technology).
\end{acknowledgments}


\appendix*

\section{Membrane deformation field}

In this appendix we derive exact analytical expressions for the displacement field for a finite-sized membrane of dimensionless radius $b^\star$ in the zeroth-order problem. 
We first calculate the normal displacement ${u_0}_z^\star$ which is dependent only on the membrane resistance toward bending.
Next, we calculate the in-plane displacements ${u_0}_x^\star$ and ${u_0}_y^\star$ which are determined by the membrane resistance toward shear.

\subsection{Bending contribution}

We consider the rescaled form of the biharmonic equation governing the evolution of a membrane resisting bending as stated by Eq.~\eqref{sigma_normal} of the main body of the paper, e.g.,
\begin{equation}
 \left(  \frac{\partial^2}{\partial {x^\star}^2} + \frac{\partial^2}{\partial {y^\star}^2}  \right)^2  {u_0}_z ^\star  =  {\sigma_0}_{zz}^\star \, , \label{sigma_normal2}	
\end{equation}
where ${\sigma_0}_{zz}^\star$ is the normal traction imposed at the planar configuration of reference as derived from the Blake tensor for a point force acting along the $x$~direction, given in the cylindrical coordinate system by~\cite{blake71}
\begin{equation}
 {\sigma_0}_{zz}^\star = - \frac{9r \left( 1+\frac{9}{16} \, a^\star \right) }{\left( 1+r^2 \right)^{5/2}}
  \cos\phi \, . \label{normalTractionBlake}
\end{equation}

Again, we consider that the membrane size is large enough for the latter expression to be valid.
For the determination of the membrane normal displacement, we use the separation of variables approach~\cite{haberman83}.
By substituting Eq.~\eqref{normalTractionBlake} into Eq.~\eqref{sigma_normal2} and transforming the resulting equation into the polar coordinate system, we readily obtain
\begin{equation}
 {u_0}^\star_{z, rrrr} + \frac{2}{r}\, {u_0}^\star_{z,rrr} + \frac{2 {u_0}^\star_{z,rr\phi\phi} - {u_0}^\star_{z, rr}}{r^2}
  + \frac{{u_0}^\star_{z,r} - 2 {u_0}^\star_{z, r\phi\phi}}{r^3} 
  +\frac{4 {u_0}^\star_{z, \phi\phi} + {u_0}^\star_{z, \phi\phi\phi\phi}}{r^4} = - \frac{9r  \left( 1+\frac{9}{16} \, a^\star \right) \cos\phi}{\left(1+r^2 \right)^{5/2}} \, .
 \label{u0z_Equation}
\end{equation}

Because of the form of the right-hand side in~\eqref{u0z_Equation}, we choose a solution of the form,
\begin{equation}
	{u_0}_z^\star = H(r) \cos\phi \, , \label{solution_uz_appendix}
\end{equation}
where the radially-symmetric function~$H$ is solution of the ordinary differential equation
\begin{equation}
 H_{,rrrr} + \frac{2H_{,rrr}}{r} - \frac{3H_{,rr}}{r^2} + \frac{3H_{,r}}{r^3} - \frac{3H}{r^4} = - \frac{9 \left( 1+\frac{9}{16} \, a^\star \right)  r}{\left(1+r^2 \right)^{5/2}} \, ,
\end{equation}
subject to the regularity conditions at $r=0$
\begin{equation}
 |H(r=0)| < \infty \, , \qquad |H_{,r}(r=0)| < \infty \, , 
\end{equation}
in addition to the boundary conditions of vanishing displacement and slope at the fixed points located at $r=b^\star$.
Specifically,
\begin{equation}
 H (r=b^\star) = 0 \, , \qquad H_{,r} (r=b^\star) = 0 \, .
\end{equation}

Under these conditions, the solution is unique and can be obtained using the algebra software package Maple as
\begin{equation}
 \begin{split}
 H(r) &= \frac{3 \left( 1+\frac{9}{16} \, a^\star \right)}{4}  \Bigg(
  2 r \ln \left( \frac{1+R}{1+\lambda} \right) - \frac{r^3}{(1+\lambda)^2} 
  + \frac{2\lambda+(\lambda-3)R+2}{R(1+\lambda)} r
  -\frac{2(R-1)}{Rr} 
  \Bigg) \, , \label{H}
 \end{split}
\end{equation}
where $R := \left(1+r^2 \right)^{1/2}$ and $\lambda := \left( 1+{b^\star}^2 \right)^{1/2}$ as defined in the main text.

By differentiating the normal displacement with respect to $x^\star$ as required by the application of the reciprocal theorem, we obtain
\begin{equation}
 {u_0}_{z,x}^\star =  H_1(r) \cos (2\phi) + H_2(r)  \, , \label{uz_x_Appendix}
\end{equation}
where the radial functions $H_1$ and $H_2$ are explicitly given by
\begin{subequations}
	\begin{align}
	 H_1 (r) &= \frac{3 \left( 1+\frac{9}{16} \, a^\star \right)}{4 R(R+1)} \left( 1-R - \frac{r^2+R-2\lambda-\lambda^2}{(1+\lambda)^2} \, r^2 \right) \, , \label{H_1} \\
		 H_2 (r) &= \frac{3\left( 1+\frac{9}{16} \, a^\star \right)}{2}  \Bigg( \frac{\lambda^2-R^2}{(1+\lambda)^2}+\ln \left( \frac{1+R}{1+\lambda} \right) \Bigg) \, .
		 \label{H_2}
	\end{align} \label{H_1H_2}
\end{subequations}

\begin{figure}
\begin{center}
\scalebox{0.85}{\input{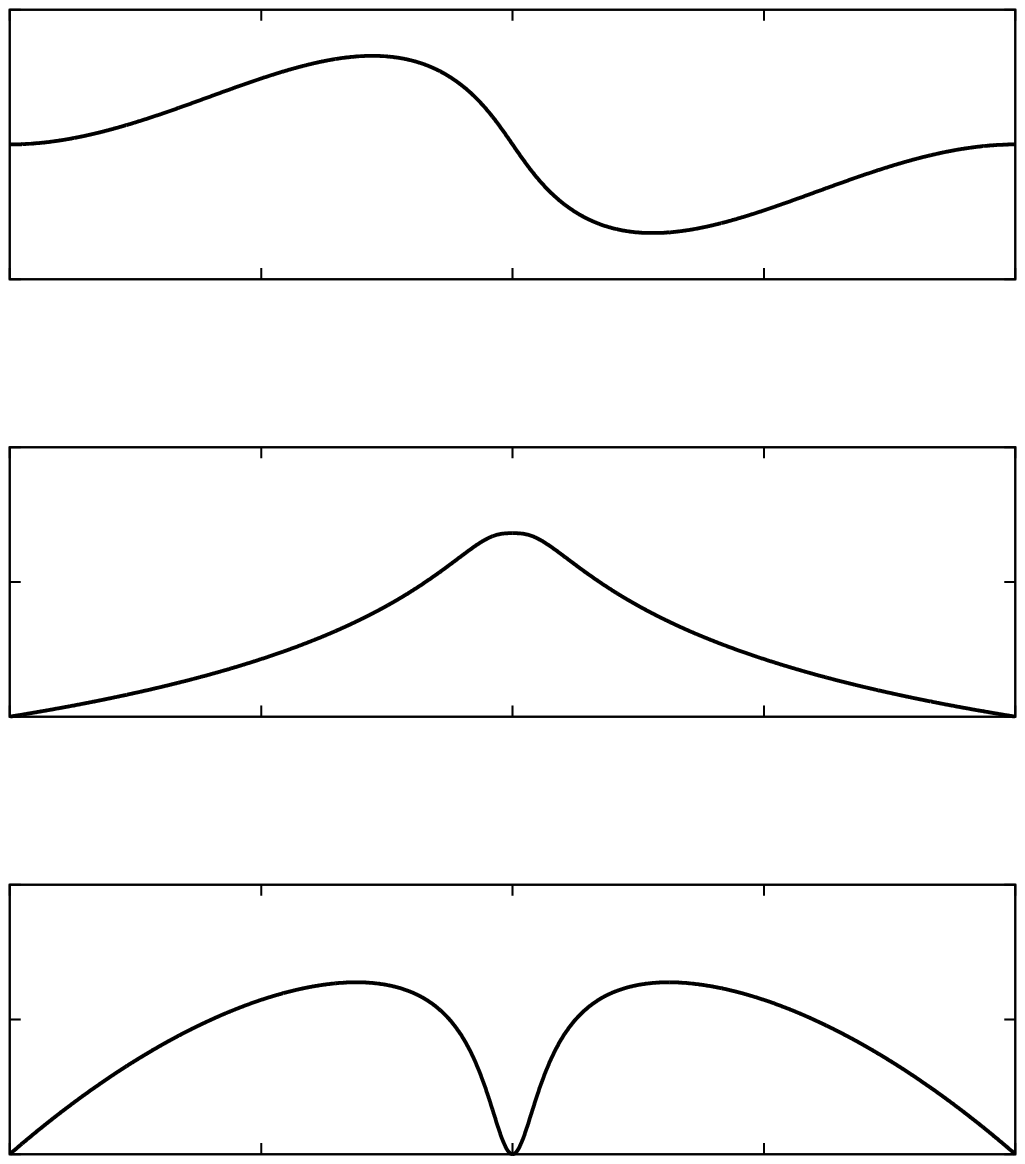}}
\end{center}
\caption{Rescaled membrane displacement in the plane of maximum deformation for $(a)$ and $(b)$ $\phi=0$, and $(c)$~$\phi=\pi/4$, as predicted theoretically for a membrane size $b^\star=20$ and Skalak ratio $C=1$.
}
\label{MemShapeElastlift}
\end{figure}

\subsection{Shear contribution}

We next consider the system of partial differential equations governing the displacement field in an elastic membrane undergoing shear deformation, stated in a condensed form by Eq.~\eqref{sigma_tangential} of the main body of the paper,
\begin{subequations}
	\begin{align}
	 -\frac{1}{3} \left( 2(1+C) \frac{\partial^2 {u_0}^\star_{x}}{\partial {x^\star}^2} 
	 + \frac{\partial^2 {u_0}^\star_{x}}{\partial {y^\star}^2} 
	 + (1+2C) \frac{\partial^2 {u_0}^\star_{y}}{\partial x^\star \partial y^\star}  \right) &=  {\sigma_0}_{xz}^\star \, , \label{sigma_tangential2} \\
		 -\frac{1}{3} \left( \frac{\partial^2 {u_0}^\star_{y}}{\partial {x^\star}^2} 
		  + 2(1+C) \frac{\partial^2 {u_0}^\star_{y}}{\partial {y^\star}^2}  
		  + (1+2C) \frac{\partial^2 {u_0}^\star_{x}}{\partial x^\star \partial y^\star} \right) &= {\sigma_0}_{yz}^\star  \, .
	\end{align}
\end{subequations}
Transforming to the polar coordinate system and using the Blake result for a point force acting along the $x$~direction, the in-plane tractions at the wall are given by \cite{blake71}
\begin{equation}
	{\sigma_0}_{xz}^\star  =  \frac{9\left( 1+\frac{9}{16} \, a^\star \right)r^2}{\left( 1 + r^2 \right)^{5/2}} \, \cos^2\phi \, , \qquad
	{\sigma_0}_{yz}^\star  =  \frac{9\left( 1+\frac{9}{16} \, a^\star \right)r^2}{\left( 1 + r^2 \right)^{5/2}} \, \cos\phi\sin\phi \, . \label{Tx_Ty}
\end{equation}

Next, considering solutions of the form
\begin{equation} \label{solution_uxuy_appendix}
	 {u_0}_x^\star (r,\phi) = A(r) \cos (2\phi) + G(r) \, , \quad\quad
	 {u_0}_y^\star (r,\phi) =  A(r) \sin (2\phi) \, , 
\end{equation}
yields the following system of differential equations in $A$ and $G$,
\begin{subequations}
	\begin{align}
	 (3+2C) \left( A_{,rr}+\frac{A_{,r}}{r}-\frac{4A}{r^2} \right)
	 +(1+2C) \left( G_{,rr}-\frac{G_{,r}}{r} \right) 
	 &= - \frac{27 \left( 1+\frac{9}{16} \, a^\star \right) r^2}{(1+r^2)^{5/2}} \, , \\
	 G_{,rr}+2(1+C) \frac{G_{,r}}{r}-A_{,rr}+2C \frac{A_{,r}}{r}+2(3+2C) \frac{A}{r^2} &= 0 \, .
	\end{align}
\end{subequations}

The solutions satisfying the regularity conditions at the origin and a vanishing displacement at the membrane extremities are unique and can be expressed as
\begin{subequations}
	 \begin{align}
	 A(r) &= \frac{9 \left( 1+\frac{9}{16} \, a^\star \right)}{8 (1+C)} \Bigg( 
	 	  \left( -\frac{2r^2}{(1+\lambda)^2}+\frac{2(R-2)}{R}+\frac{4(R-1)}{R r^2} \right) C
	 	  +1 -\frac{\lambda+2}{\lambda(\lambda+1)^2} \, r^2 - \frac{2(R-1)}{R r^2} \Bigg) \, , \label{A} \\
	 	 G(r) &= \frac{9 \left( 1+\frac{9}{16} \, a^\star \right)}{4(1+C)} \Bigg( 
	 	  (2C+3) \left( \arctanh\left(\frac{1}{\lambda}\right)-\arctanh\left(\frac{1}{R}\right)-\ln\left(\frac{r}{b^\star} \right)\right)
	 	  +\frac{1}{2C+3} \bigg( 
	 	  \frac{(2C+1)\left( 2+(2C+1)\lambda \right) }{\lambda(1+\lambda)^2} \, r^2 \notag \\
	 	  &-\frac{4C^2(\lambda-1)}{\lambda+1} + \frac{2C \left( 3R-\lambda-2R\lambda \right)}{R\lambda}
	 	  -\frac{3}{R} + \frac{5-(\lambda-2)\lambda}{\lambda(\lambda+1)} 
	 	  \bigg)
	 	  \Bigg) \, . \label{G}
	\end{align} \label{AG}
\end{subequations}

By taking the derivatives of the in-plane displacements with respect to $x^\star$, as required by the application of the reciprocal theorem, we obtain
\begin{equation}\label{ux_x_uy_x_Appendix}
	 {u_0}^\star_{x,x} = \cos\phi \left( K_1(r) \cos^2\phi + K_2(r) \right)\, , \quad\quad
	 {u_0}^\star_{y,x} = \sin\phi \left( W_1(r) \cos^2\phi + W_2(r) \right) \, ,
\end{equation}
where we have defined
\begin{subequations}\label{K1K2W1W2}
	\begin{align}
	 K_1 (r) &= -\frac{9 \left( 1+\frac{9}{16} \, a^\star \right)}{2 (1+C)R^3 r} \bigg( \big( (1+2C)R-6C \big)r^2+5-14C+(10C-3)R + \frac{4(2C-1)(R-1)}{r^2} \bigg)  \, , \label{K1} \\
	 	  K_2 (r) &= \frac{9(2C-1) \left( 1+\frac{9}{16} \, a^\star \right)}{4	 (1+C)} \Bigg( \frac{r}{(\lambda+1)^2 (2C+3)} \left(1+2C+\frac{2}{\lambda}\right)
	 	  +\frac{1}{rR} \left( R-4+\frac{6(R-1)}{r^2} \right)
	 	  	 	  \Bigg) \, , \\
	 	 W_1 (r) &= - \frac{9 \left( 1+\frac{9}{16} \, a^\star \right)}{2 (1+C)Rr} \Bigg( 2C\left( R-3+\frac{4(R-1)}{r^2} \right) + \frac{1}{R} \left( r^2+\frac{5-3R}{R} - \frac{4(R-1)}{r^2 R} \right) \Bigg) \, , \\
	 	 W_2 (r) &= \frac{9 \left( 1+\frac{9}{16} \, a^\star \right)}{4 (1+C)r} \Bigg( 
	 	  2C \left( -\frac{r^2}{(\lambda+1)^2}+\frac{R-2}{R}+\frac{2(R-1)}{r^2 R} \right)
	 	  +1-\frac{2+\lambda}{\lambda(\lambda+1)^2} \, r^2 - \frac{2(R-1)}{r^2 R} \Bigg) \, . \label{W2}
	\end{align} 
\end{subequations}

Fig.~\ref{MemShapeElastlift} illustrates the variation of the displacement fields along the membrane as predicted theoretically in Eq.~\eqref{solution_uz_appendix} for the normal displacement, and Eq.~\eqref{solution_uxuy_appendix} for the in-plane displacements.
Here the membrane size is set $b^\star=20$ and the Skalak ratio $C=1$.
The displacements are shown in their plane of maximum deformation corresponding to $\phi=0$ for ${u_0}^\star_z$ and ${u_0}^\star_x$ (Fig.~\ref{MemShapeElastlift}~$(a)$ and $(b)$), and to the plane $\phi=\pi/4$ for ${u_0}^\star_y$ (Fig.~\ref{MemShapeElastlift}~$(c)$).
The normal displacement is found to be about one order of magnitude larger that the lateral displacements.
This is in accord with the calculations of the lift force where the effect of membrane resistance toward bending is found to be more significant compared to that of shear.

%

\end{document}